\documentclass[aps,showpacs,prb,floatfix,twocolumn]{revtex4}
\usepackage{amsmath,amssymb,graphicx,bm,epsfig}
\usepackage{epstopdf}

\newcommand{\vk}{{\mathbf{k}}}

\begin{document}

\title{Temperature-dependent Fermi surface evolution 
in heavy fermion CeIrIn$_5$}

\author{Hong Chul Choi$^1$, B. I. Min$^1$, J. H. Shim$^{1,2}$, 
K. Haule$^3$, and G. Kotliar$^3$}
\affiliation{$^1$Department of Physics, 
Pohang University of Science and Technology, Pohang 790-784, Korea}
\affiliation{$^2$Department of Chemistry, 
Pohang University of Science and Technology, Pohang 790-784, Korea}
\affiliation{$^3$Department of Physics, 
Rutgers University, Piscataway, NJ 08854, USA}

\pacs{71.18.+y,71.27.+a,72.15.Qm}
\date{\today}
\maketitle
{\bf
In Cerium-based heavy electron materials, 
the $4f$ electron's magnetic moments bind 
to the itinerant quasiparticles 
to form composite {\it heavy} quasiparticles at low temperature ($T$). The volume 
of the Fermi surface (FS) in the Brillouin zone incorporates the moments
to produce a "large FS" due to the Luttinger theorem.
When the $f$ electrons are localized free moments, a "small FS" is induced
since it contains only broad bands of conduction $spd$ electrons. 
We have addressed theoretically the evolution 
of the heavy fermion FS as a function of $T$, using 
a first principles dynamical 
mean-field theory (DMFT) approach combined with density functional 
theory (DFT+DMFT). 
We focus on the archetypical heavy electrons 
in CeIrIn$_5$, which is believed to be near a quantum critical point. 
Upon cooling, both the quantum oscillation frequencies and 
cyclotron masses show logarithmic scaling behavior ($\sim$ ln$(T_0/T)$)
with different characteristic 
temperatures $T_0$ = 130 and 50 K, respectively. The enlargement of the 
electron FS's at low $T$ is accompanied by 
topological changes around T = 10 $\sim$ 50 K. The resistivity 
coherence peak observed at $T \simeq 50$ K is the result of 
the competition between the binding of 
incoherent $4f$ electrons to the $spd$ conduction electrons at Fermi level ($E_F$) and 
the formation of coherent $4f$ electrons. 
}

The FS volume has been
a sensitive probe of the character, localized or itinerant, of the heavy 
fermion system.\cite{Settai07} 
Intensive efforts have been devoted to the study of the 
quantum phase transition leading from a small to large FS at 
strictly zero temperature.
While the FS, as a surface of discontinuity in the momentum 
distribution function, is sharply defined only at zero temperature, 
experimental probes such as the angle-resolved 
photoemission spectra (ARPES) and magnetic quantum 
oscillation experiments such as de Haas-van Alphen 
(dHvA) or Shubnikov-de Hass experiments identify the region of momentum 
space where zero energy fermionic excitations exist at finite 
temperature. 
ARPES experiment directly observes the FS in the momentum space.
But, high resolution is required to determine the FS size. 
The quantum oscillation experiments measure the 
precise value of the FS area in a specific plane 
by probing the oscillation frequencies of 
magnetization as a function of the applied magnetic field. 
The quantum oscillation frequency($\mathrm{F}$), the so-called dHvA frequency,
is proportional to the extremal cross-sectional 
area $S_F$ of the FS ($\mathrm{F} =\hbar S_F/2\pi e$).
The quantum oscillation experiments also provide the information on the 
cyclotron effective electron mass 
$m^* (= (\hbar^2 / 2 \pi)  \partial S_F / \partial \omega$) and
the geometry of the FS's. 

The band structure calculation is a complementary tool to 
the quantum oscillation experiment to analyze the complicated FS 
of the multiple band system. 
Quantum oscillation frequencies of heavy fermion materials, such as 
CeCu$_6$, UPt$_3$, and Ce(Ir,Co)In$_5$, are explained well by
conventional band calculations 
because the itinerant $4f$ electrons 
behave as conduction electrons near $E_F$. 
Although the geometry and volume of FS's are 
well explained by the DFT band calculation, the detected cyclotron mass 
$m^*$ is much larger than the corresponding 
DFT band mass $m_b$,\cite{Elgazzar,Haga,Settai01,Onuki} 
because the DFT calculation can not describe the correlated $4f$ 
electronic states correctly.  When the $4f$ electrons are 
localized in the antiferromagnetic (AFM) compounds, 
such as CeRhIn$_5$, CeIn$_3$, 
CeRh$_2$Si$_2$, the $4f$-localized band model is more applicable 
to the description of the quantum oscillation experiments.\cite{Elgazzar,Onuki}
The $4f$-localized band model can be performed 
by treating the $4f$ electrons as core within the DFT (open-core DFT)
band calculation,\cite{Elgazzar} or by employing the DFT+$U$ band
method ($U$: on-site Coulomb interaction).\cite{Wang}

Ce$T$In$_5$ ($T$ = Co, Rh, and Ir) has been a prototypical system to study
the crossover behavior between the itinerant and localized $4f$ electrons. 
CeCoIn$_5$\cite{Petrovic} and CeIrIn$_5$\cite{Petrovic2} have itinerant
$4f$ electrons and superconducting ground states at low $T$.
On the other hand, CeRhIn$_5$ has localized $4f$ electrons and 
the AFM ground state. 
The measured dHvA frequency of each compound
identifies the nature of Ce $4f$ electrons, whether they are 
itinerant or localized.
CeCoIn$_5$\cite{Elgazzar,Hall,Settai01} and CeIrIn$_5$\cite{Elgazzar,Haga} have 
enlarged electron FS's due to the contribution of itinerant $4f$ electrons, 
while CeRhIn$_5$ has similar geometry of FS's
but smaller size of FS's.\cite{Elgazzar,Onuki} 
For CeRhIn$_5$, pressure-induced superconductivity was 
observed for $P > 1.63$ GPa,\cite{Hegger}
and the drastic change in the FS was detected at a critical pressure of
$P_c \simeq 2.35$ GPa.\cite{Shishido}
On the other hand, CeRh$_{1-x}$Co$_{x}$In$_5$ 
shows the the doping-dependent reconstruction of FS deep inside the magnetically ordered state,\cite{Goh} 
away from the quantum critical point.
The $T$-dependent evolution between itinerant and localized electrons 
also has been described by the phenomenological two-fluid model, 
where the universal scaling behavior can be applied to various physical 
properties of the heavy fermion compounds.\cite{Pines04,Pines08,Yifeng}

In this Letter, we have addressed the $T$-dependent crossover 
from localized to itinerant $4f$ electrons in CeIrIn$_5$,
and investigated its effects on the FS properties 
and electrical resistivity.
The charge self-consistent version of DFT+DMFT approach,\cite{dmft} as implemented in Ref.17, is
based on the full-potential linearized augmented plane-wave (FP-LAPW) band method.\cite{wien2k}
The correlated $4f$ electrons are treated dynamically by the DMFT local self-energy, 
while all other delocalized $spd$ electrons are treated on the DFT level.
The local self-energy matrix $\Sigma(\omega)$ is calculated from
the corresponding impurity problem, in which  
full atomic interaction matrix is taken into account.\cite{Cowan} 
To solve the impurity problem, we use both the vertex corrected 
one-crossing approximation\cite{dmft} 
and the continuous time quantum Monte-Carlo method.\cite{Haule,Werner} 
  
The main difference between low and high $T$ spectral functions
in the DFT+DMFT calculation
is the existence of $4f$ bands near $E_F$, 
as shown in Supplementary Fig.~1.  
Ce $4f$ bands at high $T$ are 
absent near $E_F$, and their spectral weights are distributed
into the lower and upper Hubbard bands.
The spectral function near $E_F$ can be well described 
by the quasiparticle band structures
of other $spd$ electrons although there is a small scattering rate  
due to the hybridization
between the conduction electrons at $E_F$ and the incoherent Ce $4f$ electrons 
in the Hubbard bands.
As decreasing $T$, the spectral weight of the renormalized Ce $4f$ bands 
is increased continuously (see Supplementary Movie 1.).
The hybridization of the $4f$ and other $spd$ bands
produces very massive almost flat quasiparticle band structures near $E_F$.
These flat bands emerge as the narrow Kondo peak at $E_F$ 
in the photoemission spectra.\cite{Shim}

\begin{figure}[tb]
\includegraphics[width=1.0\linewidth]{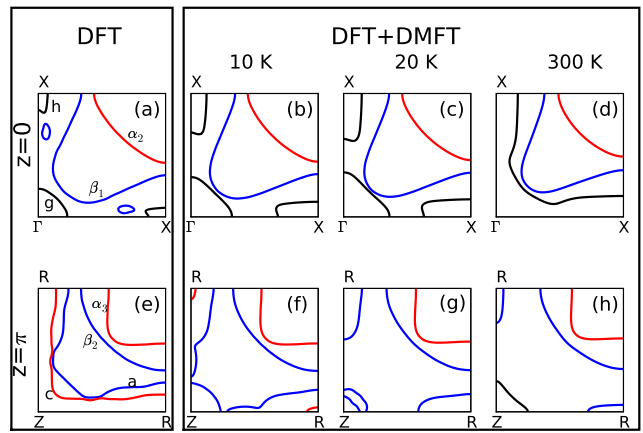}
\caption{
{\bf
The $T$-dependent FS evolution in the DFT+DMFT calculation.
}
The FS's are extracted from the DFT+DMFT
quasiparticle band structures at 10 K(b,f), 20 K(c,g) and 300 K(d,h).
For comparison, the FS's obtained from the DFT band calculation 
are also provided (a,e).
Because the main FS's in CeIrIn$_5$ are nearly cylindrical
due to the quasi-2D nature of its crystal structure,
the FS's only on the $z=0$ and $z=\pi$ planes are shown. 
The FS's on these planes are identified from 
the dHvA frequencies because the symmetric plane provides 
the extremal cross-section of the FS.
There are two main cylindrical electron FS's 
represented by $\alpha_i$ and $\beta_i$ branches 
observed in the dHvA experiment (see (a) and (e)). 
Those branches are identified at all temperature range.
On the other hand, the FS's denoted as 
$g$, $h$ (hole FS's) and $a$, $c$ (electron FS's)
in the DFT calculation manifest topological changes with 
varying $T$ in the DFT+DMFT calculation.
Note that $g$, $h$, $a$, $c$ branches were not identified clearly 
in the dHvA experiments.
Color represents the different band index.
}
\label{fig1}
\end{figure}

The $T$-dependent FS has been extracted from the quasiparticle band structures.
At low $T$, the FS's of the DFT+DMFT calculation are very similar 
to those of the DFT calculation, as shown in Fig. 1. 
Upon heating, the Ce $4f$ electrons become localized 
and their contribution to $E_F$ is suppressed.
Accordingly, the areas of electron FS's ($\alpha_i$ and $\beta_i$)
are continuously decreased.
In contrast, there occur rather big changes in other FS's areas
upon heating.
The areas of the $g$ and $h$ hole FS's on the $z=0$ plane grow
and merge into one closed electron FS.
The $a$ electron FS identified at $T=10$ K on the $z=\pi$ plane is 
divided at high $T$, and so new hole FS's appear 
near Z and R symmetry points.
The continuous $T$-dependent evolution of the FS is provided in Supplementary Movie 2.
By integrating the volume of electron FS's, the occupancy of the conduction electrons
has been counted. It shows the continuous change from 3 to 4 electrons 
as temperature is decreased, which reflects
the participation of one Ce $4f$ electron in the bonding.

\begin{figure*}[tb]
\includegraphics[width=1.0\linewidth]{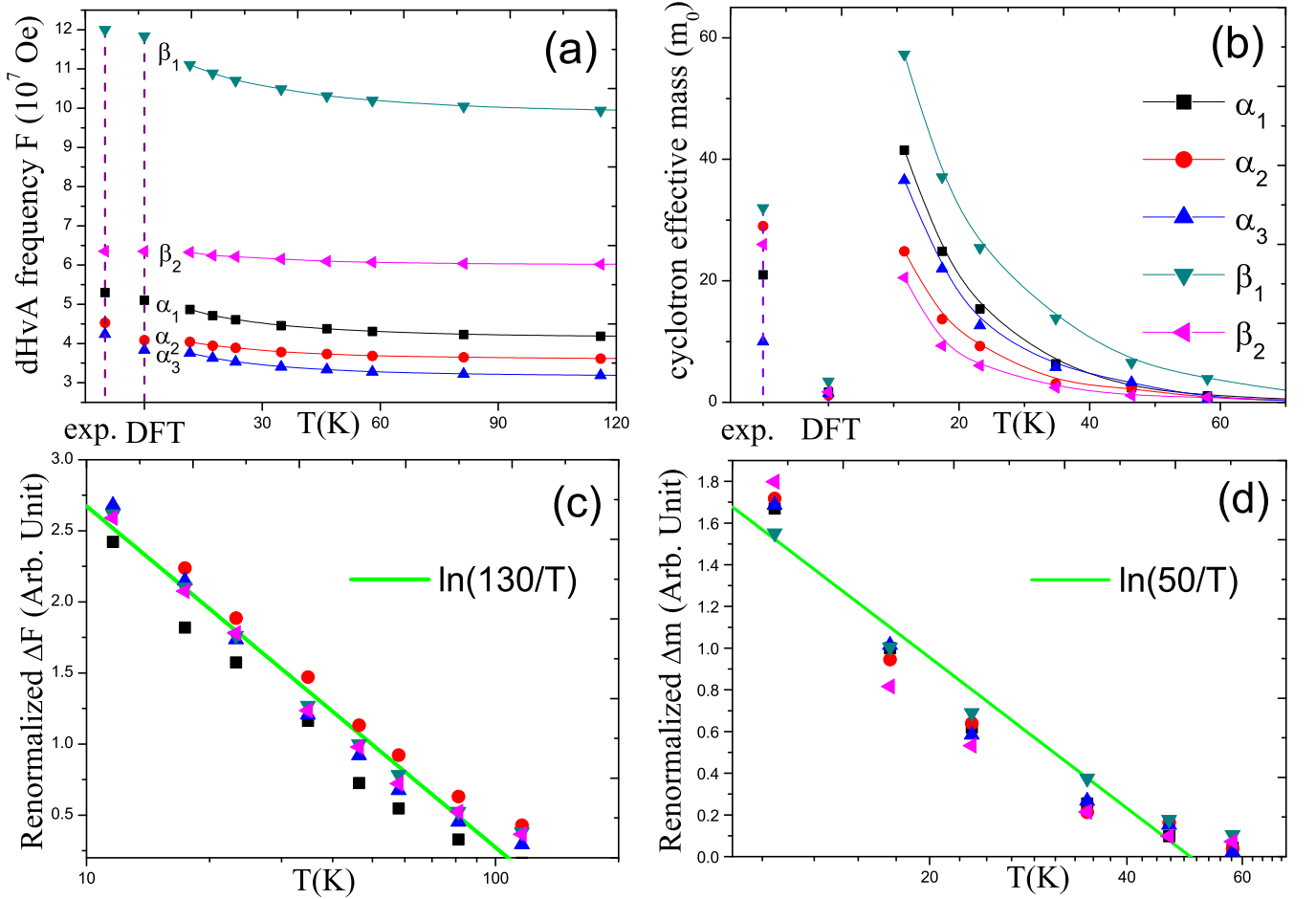}
\caption{
{\bf
The $T$-dependent dHvA frequencies ($\mathrm{F}$) and cyclotron effective masses ($m^*$).
}
The dHvA frequencies (a) and effective masses (b) of $\alpha_i$ and 
$\beta_i$ branches 
are obtained from the DFT+DMFT method, and compared with those 
from the DFT method and experiments (exp).
The corresponding FS's for each branch are provided in Fig. 1(a,e), 
except $\alpha_1$
that corresponds to the maximum frequency among $\alpha_i$ branches 
and is located between the $z=0$ and $z=\pi$ planes.
At high $T$, the cyclotron masses are very small, 
ranging from 0.4 to 0.7 $m_0$ 
($m_0$: bare electron mass) for $\alpha_i$ and $\beta_i$ branches. 
Such small cyclotron masses are also reproduced in the $4f$ 
open-core DFT calculation, 
in which only dispersive $spd$ bands are crossing $E_F$.
The low $T$ dHvA frequencies from the DFT+DMFT method are consistent 
with the results of DFT method in which the $4f$ electrons are considered 
as itinerant type. 
(c) The renormalized $\Delta \mathrm{F}_i$ of each branch shows the scaling
behavior of ln$(T_0/T)$ with the characteristic $T$ of $T_0^f \sim 130$ K.  
(d) All the renormalized $\Delta m^*$'s show the similar scaling behavior, but with $T_0^m \sim 50$ K.
}
\label{fig2}
\end{figure*}

Because the area of the FS is directly related to the size of the $4f$ 
electron contribution to $E_F$,
we have investigated the $T$-dependent dHvA frequencies, as shown in Fig. 2(a).
At high $T$, the dHvA frequencies are well consistent with 
those from the Ce $4f$ open-core DFT calculation.
With decreasing $T$, they show the continuous increase with the participation 
of $4f$ electrons to $E_F$ and follow the scaling behavior of ln$(T_0/T)$, 
as shown in Fig. 2(c). All the branches
show the same characteristic temperature $T_0^f \sim$ 130 K.  
This behavior is consistent with the increase of the number 
of conduction electrons with decreasing $T$.

The cyclotron mass corresponds to the effective mass of the carriers 
at the specific FS.	
As shown in Fig. 2(b) and (d), the calculated cyclotron masses also 
increase upon cooling, and follow a similar scaling behavior 
of $\sim$ ln$(T_0/T)$ with $T_0^m \sim$ 50 K.
The cyclotron masses are also well fitted by the description of two fluid model by Yang {\it et al.}\cite{Pines08}: 
$(1-T/T_0)^{3/2}[1+{\rm ln}(T_0/T)]$ with same $T_0^m \sim 50$ K.
Interestingly, $T_0^m$ is 
coincident with the coherent temperature $T^*$ of Ce $4f$ states,\cite{Shim}
but clearly different from $T_0^f$.
This feature reveals the $4f$ electrons start to participate in bonding through the hybridization 
with $spd$ electrons at the temperature scale $T_0^f$, which is higher than the temperature $T_0^m$ at which the coherent heavy fermion electronic states are formed.
These results are reminiscent of recent experiment, which shows the occurrence of FS reconstruction much earlier than the quantum critical transition.\cite{Goh}
Note that the above scaling law is consistent with the two-fluid model,\cite{Pines08}
in which the coherent $4f$ bands start to grow below $T^*$.

All the calculated cyclotron masses at $T=10$ K  
seem to be overestimated with respect to the experimental values\cite{Haga} 
roughly by a factor of two.
It is well known that the value of cyclotron mass has a substantial 
dependence on the applied magnetic field.\cite{Settai01}
In the presence of the magnetic field,
the effective mass can be reduced by the change of the 
hybridization,
even without much change of the FS geometry.\cite{Lawrence} 
If one considers the high magnetic field in experiments, 
the calculated cyclotron masses would be consistent 
with experimental values at low $T$.

\begin{figure}[tb]
\includegraphics[width=1.0\linewidth]{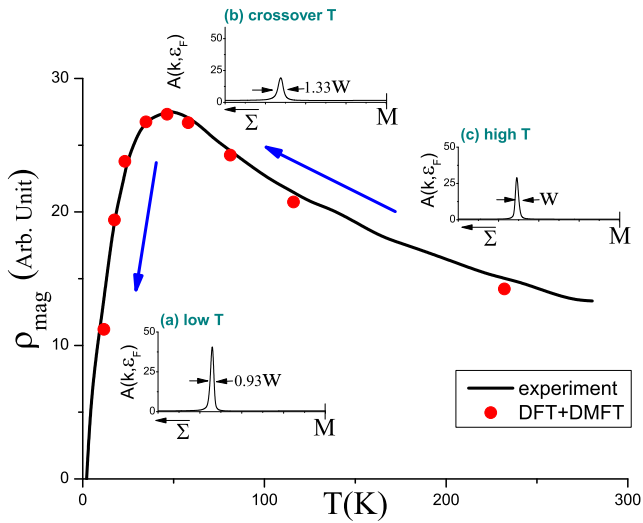}  
\caption{
{\bf
The magnetic part ($4f$ electron contribution) of the resistivity as a function of $T$.
}
The experimental electrical resistivity is obtained by subtracting 
the resistivity of LaIrIn$_5$ from that of CeIrIn$_5$.\cite{Onuki}
Inset Fig. (a), (b), and (c) show the broadening changes of spectral weights 
at $E_F$ at low (10K), crossover (50 K), and high temperature (1000 K), respectively.
$\Sigma$ means the direction from $M$ to $\Gamma$ in momentum space.
}
\label{fig3}
\end{figure}

The continuous change of FS properties with $T$ variation is 
deeply related to the transport properties. 
Figure~ 3 (a) provides the calculated resistivity for CeIrIn$_5$
as a function of $T$, which is compared to the experimental 
electrical resistivity.
The electrical resistivity is calculated
using the real part of the dc conductivity ($\sigma$)\cite{Haule_w2k} 
based on the DFT+DMFT spectral function near $E_F$:
$\sigma^{\mu\upsilon}=\frac{\pi e^2}{V} \sum_{\vk} 
{ \int d\omega \left(  -\dfrac{df}{d\omega} \right)  
\mathrm{Tr} \left[ A(\vk,\omega)v^{\vk\mu}A(\vk,\omega)v^{\vk\nu} \right] }$.
Here $\mu$ and $\nu$ represent spatial coordinates. $V$, $f(\omega)$, and $v$ 
are the primitive volume, the Fermi Dirac distribution function, 
and the velocity, respectively. 
The calculated resistivities from low to high $T$
are in good agreement with the experimental resistivity.
At high $T$, the electronic carriers from dispersive $spd$ bands become more and more decoupled 
from localized electrons in the $4f$ shell, hence the carriers are scattered less at very high $T$. 
Upon cooling, the hybridization among local moments and $spd$ carriers increases 
while the $4f$ electrons remain very incoherent above 50 K, 
causing enhanced scattering mechanism for electric carriers.  
Below the scale $T_0^m$, the electrons in the $4f$ shell also gain coherence which substantially suppresses resistivity. Therefore, the maximum resistivity is observed near 50 K.
Inset Fig. 3(a), (b), and (c) show the broadening of spectral weight at $E_{F}$,
calculated at low, crossover, and high $T$, respectively.
The broadening corresponds to the scattering rate at the specific {\bf k}-point.
It is noted that the spreading of the spectral weight 
at crossover is wider than that at high or low $T$.
This finding confirms that the DFT+DMFT calculation
describes well the crossover behavior of Ce $4f$ electrons
with one $T_0^f$ ($\sim 130$ K) for the participation of $4f$ electrons in the conduction
and another $T_0^m$ ($\sim 50$ K) for the formation of coherent heavy electron $4f$ bands.

We have examined the evolution of the heavy fermion state using electronic structure methods.
As in the two fluid phenomenology,\cite{Pines04} the experimental
studies of other heavy fermion systems\cite{Gegenwart} as well as the slave boson studies,\cite{Burdin, Burdin2}
the crossover from the high $T$ regime, where moments and quasiparticles coexist,
to the low $T$ Fermi liquid heavy fermion state, has a rich structure characterized by multiple energy scales.
We have found that it is characterized by multiple scales 
which have a clear correspondence with physical observables. 
$T_0^f$ is the onset of the sharp crossover where the small FS begins distorting towards the low $T$ FS. 
At a lower $T_0^m$, composite quasiparticles formed from $f-$moments and conduction electrons emerge, and this is signaled by a maximum of the resistivity.   
By that point, the FS has reached a shape which is closer to its zero temperature final value, but the material is not yet a Fermi liquid, which is only reached at a much lower temperature $T_{FL}$. 
We can only put bounds for this quantity as being lower than 10 K for the 115 material.

The theory can be tested using several techniques such as ARPES, Compton scattering and scanning tunnelling microscopy, which have been developed as powerful tools for exploring the evolution of the electronic structure and are currently under way.\cite{Delinger}  
Our theory predicts that both $T_0^m$ and $T_0^f$ increase as a function of pressure in  the CeIrIn$_5$ material.
More generally, it would be interesting to follow these scales as a function of control parameters such as pressure and composition, to investigate the behavior of $T_0^m$ and $T_0^f$ in related materials which can be driven to a quantum critical point.

\begin{acknowledgments}
We acknowledge useful discussions with Tuson Park.
This work was supported by the NRF (No. 2009-0079947, 2010-0006484, 2010-0026762),
WCU through KOSEF (No. R32-2008-000-10180-0),
and by the POSTECH BK21 Physics Division.
K. Haule was supported by Grant NSF NFS DMR-0746395 and Alfred P. Sloan fellowship. 
G. Kotliar was supported by NSF DMR-0906943
\end{acknowledgments}

\end{document}


\title{Temperature dependent Fermi surface evolution 
in heavy fermion CeIrIn$_5$
-Supplementary Information-
}

\author{Hong Chul Choi$^1$, B. I. Min$^1$, J. H. Shim$^{1,2}$, K. Haule$^3$, and G. Kotliar$^3$}
\affiliation{$^1$Department of Physics, 
Pohang University of Science and Technology, Pohang 790-784, Korea}
\affiliation{$^2$Department of Chemistry, 
Pohang University of Science and Technology, Pohang 790-784, Korea}
\affiliation{$^3$Department of Physics, 
Rutgers University, Piscataway, NJ 08854, USA}	
	
\pacs{71.18.+y,71.27.+a,72.15.Qm}
\date{\today}
\maketitle

\section{Spectral function ($A(k,\omega)$)}
\begin{figure}[tb]
\begin{center}$
\begin{array}{cc}
\includegraphics[width=1.0\linewidth]{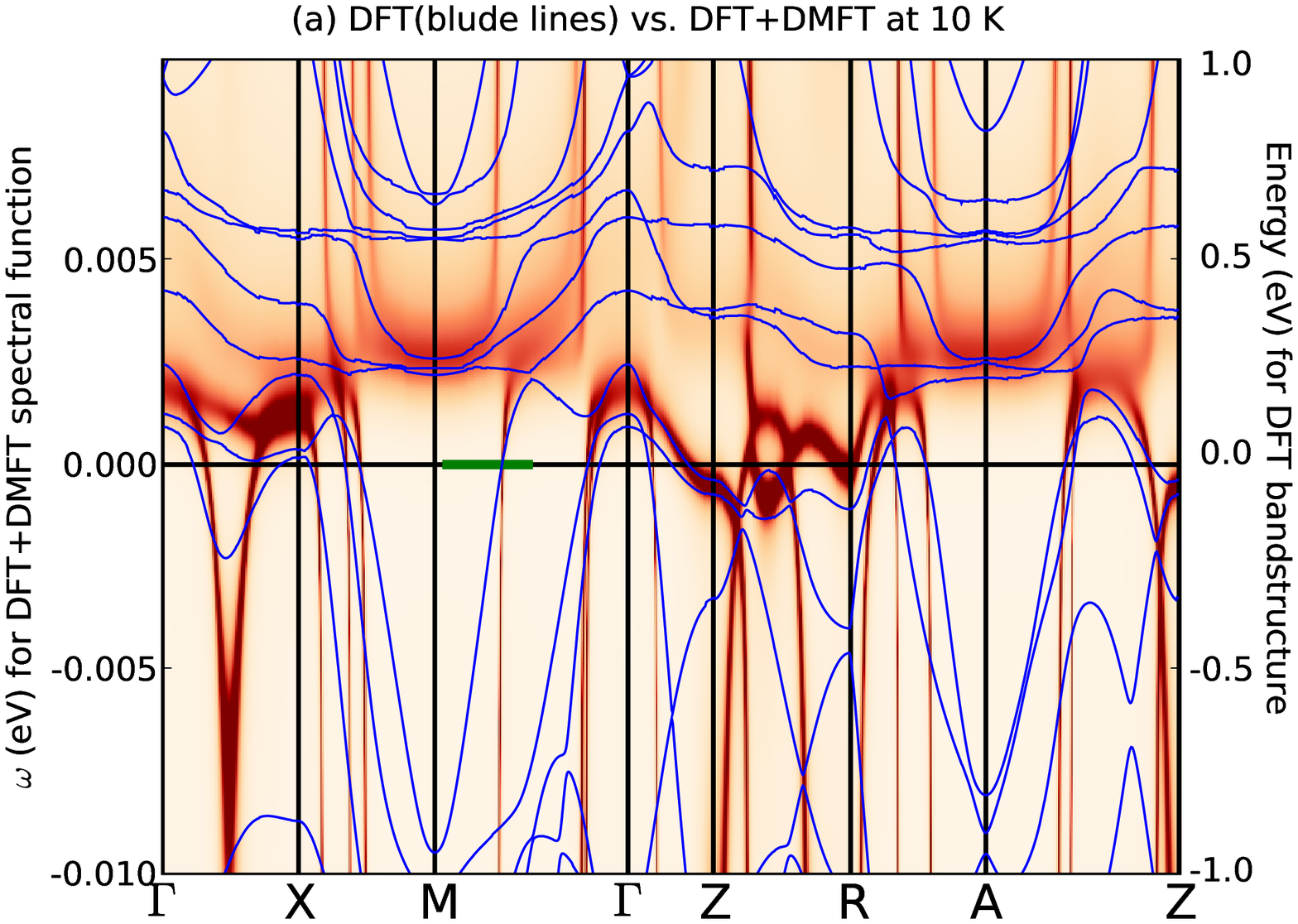}\\
\includegraphics[width=1.0\linewidth]{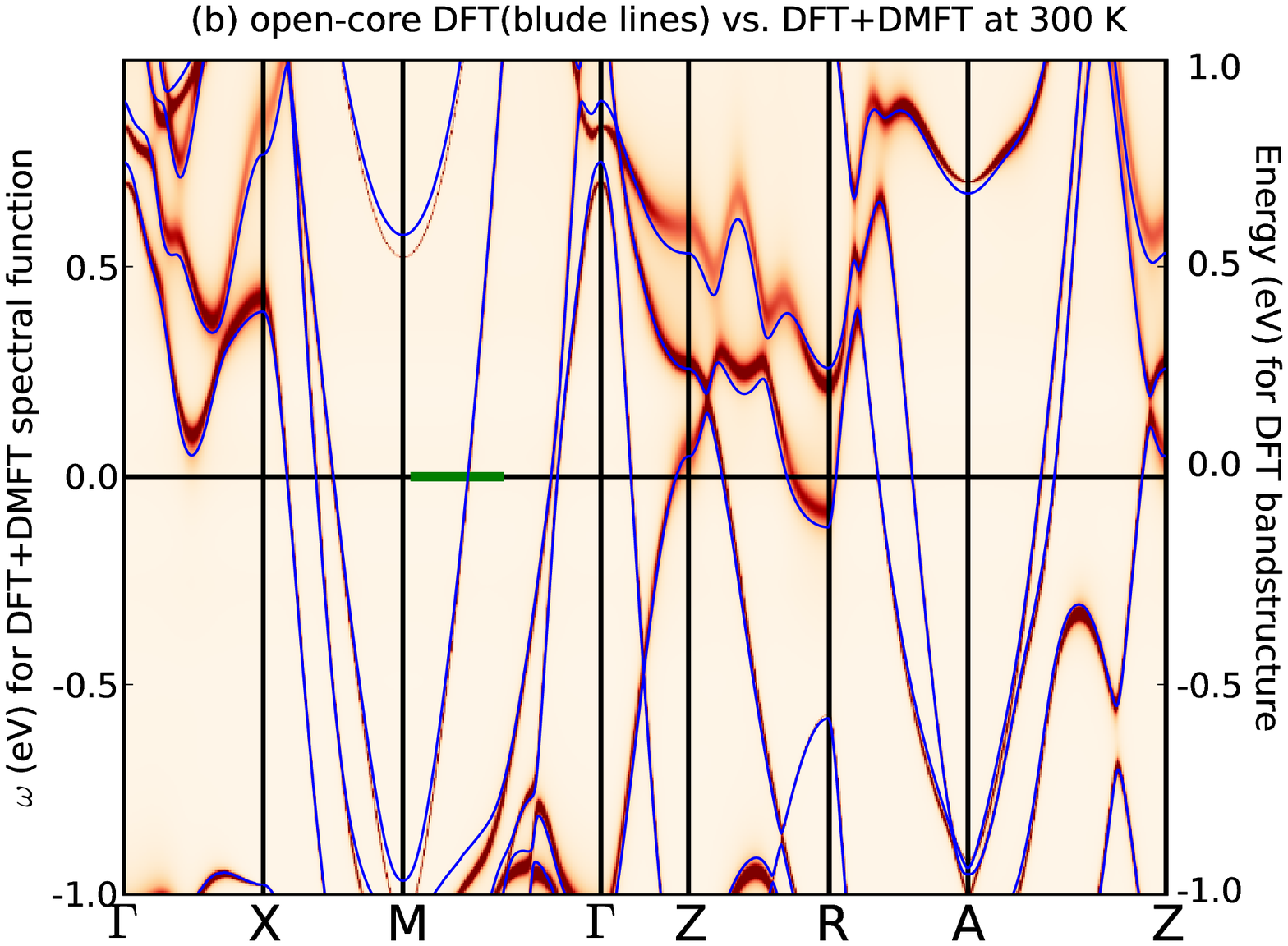}
\end{array}$
\end{center}
\caption{
{\bf
The DFT band structures and DFT+DMFT spectral functions at high and low $T$.
}
The DFT+DMFT spectral functions calculated at $T$ = 10 K (a) and 300 K (b)  
along the symmetry lines are provided. 
The spectral function $A(k,\omega)$ is obtained 
from $A(k,\omega)= -\frac{1}{\pi} \mathrm{Im} G(k,\omega)$.
In (a), the DFT band structures, in which Ce $4f$ electrons 
are treated as valence electrons (delocalized limit),
are drawn with blue lines.
In (b), the $4f$ open-core DFT band structures,
in which Ce $4f$ electrons are treated as core electrons (localized limit),
are drawn with blue lines.
The green lines along M and $\Gamma$ indicate the {\bf k}-point
paths used in Inset Fig.~3 (a), (b), and (c) of the Letter.
}
\label{fig1}
\end{figure}
 The main difference between the conventional DFT band structure and 
the Ce $4f$-localized (open-core) DFT band structure
is the existence of $4f$ bands near $E_F$, 
as shown in Supplementary Fig.~1 (blue lines). 
In the DFT band structure in Supplementary Fig.~1(a), 
one can notice the spin-orbit (SO) split $j$ = 5/2 and 7/2 states of 
Ce $4f$ flat bands, which are located around 
0.3 and 0.6 eV, respectively (clearly distinguishable at M and A).
They are hybridized with other conduction ($spd$) bands to form 
dispersive bands near $E_F$. 
Three bands containing Ce $4f$ ($j$ = 5/2) states 
are crossing $E_F$, and the occupancy of these bands is counted 
to four electrons: three $spd$ electrons and one Ce $4f$ electron.
On the other hand, in the $4f$ open-core DFT band structure in Supplementary Fig.~1(b), 
there is no contribution from Ce $4f$ electrons 
and only dispersive $spd$ bands are observed. 
The occupancy of these bands is counted to three electrons 
due to the absence of one $4f$ electron.

We have also provided the DFT+DMFT spectral functions 
obtained at low ($T=10$ K) and high ($T=300$ K) temperature. 
It is seen in Supplementary Fig.~1(b) that Ce $4f$ bands at high $T$ are completely 
removed from $E_F$, and their spectral weights are redistributed
into lower and upper Hubbard bands at -2.5 eV and +3 eV (not shown here).
The spectral function near $E_F$ can be well described 
by the quasi-particle band structures of 
$spd$ electrons only,
 although there is small scattering rate  due to the hybridization
between conduction electrons at $E_F$ and incoherent Ce $4f$ electrons 
at the Hubbard bands.
Noteworthy is that the quasi-particle $spd$ bands near $E_F$
are very  similar to those of the $4f$ open-core DFT band calculation. 

At low $T$ in Supplementary Fig.~1 (a), the DFT+DMFT spectral functions near $E_F$ 
have quasi-particle band structures of Ce $4f$ ($j$=5/2) states 
that are hybridized with $spd$ states. 
The bandwidth of Ce $4f$ ($j$=5/2) states is about 3 meV, which 
is $\sim 1/100$ smaller than that obtained in the DFT band calculation.
That is why they are shown as straight lines in this narrow energy range. 
The renormalized Ce $4f$ bands are hybridized with other $spd$ bands 
to produce flat quasi-particle band structures, as shown in Supplementary Fig.~1(a).
These flat bands emerge as the narrow Kondo peak at $E_F$
in the photoemission spectra. 
Because Ce $4f$ ($j$=5/2) states are simply renormalized 
with respect to $E_F$,
the quasi-particle bands reproduce the FS's of the DFT band calculation.

\section{the continuous $T$-dependent images of the spectral functions}
The images for moving pictures (Movies 1 and 2) are provided 
for the additional Supplementary information.
Movie 1 shows the $T$-dependent variation of $A(k,\omega)$ 
on the {\bf k}-points between $\Gamma$ and $X$ from $T=10$ to 1000 K.
The Movie 2 demonstrates the $T$-dependent variation of FS's on the $z=0$ plane from $T=10$ to 1000 K.
